\documentclass[%
reprint,
superscriptaddress,
 amsmath,amssymb,
 aps,
floatfix,
]{revtex4-1}

\usepackage{graphicx}
\usepackage{dcolumn}
\usepackage{bm}
\usepackage{color}
\usepackage{slashed}

\newcommand{\iu}{{i\mkern1mu}}

\begin{document}


\title{Reduced Density Matrices and Moduli of Many-Body Eigenstates}

\author{Chaoming Song}
\email{c.song@miami.edu}
\affiliation{%
Department of Physics, University of Miami, Coral Gables, Florida 33146, USA}%




\begin{abstract}
Many-body wavefunctions usually lie in high-dimensional Hilbert spaces. However, physically relevant states, i.e, the eigenstates of the Schr\"odinger equation are rare. For many-body systems involving only pairwise interactions, these eigenstates form a low-dimensional subspace of the entire Hilbert space. The geometry of this subspace, which we call the eigenstate moduli problem is parameterized by a set of 2-particle Hamiltonian. This problem is closely related to the $N$-representability conditions for 2-reduced density matrices, a long-standing challenge for quantum many-body systems. Despite its importance, the eigenstate moduli problem remains largely unexplored in the literature. In this Letter, we propose a comprehensive approach to this problem. We discover an explicit set of algebraic equations that fully determine the eigenstate spaces of $m$-interaction systems as projective varieties, which in turn determine the geometry of the spaces for representable reduced density matrices. We investigate the geometrical structure of these spaces, and validate our results numerically using the exact diagonalization method. Finally, we generalize our approach to the moduli problem of the arbitrary family of Hamiltonians parameterized by a set of real variables. 
\end{abstract}

\maketitle



Finding eigenstates is one of the central challenges for many-body quantum systems \cite{coleman2015introduction}. Except for a few limited cases \cite{sutherland2004beautiful}, the exact solutions of most many-body systems are unknown. Many applications rely on computational methods such as the density functional \cite{dreizler2012density}, quantum Monte Carlo \cite{becca2017quantum}, and other variational methods \cite{orus2019tensor}. On the other hand, most interesting systems involve only pairwise interactions ($2$-interaction), implying that their eigenstates form a small subspace of the many-body Hilbert space. To the best of our knowledge, the geometrical structure of this subspace, as we named it the eigenstate moduli problem, has been largely unexplored in the literature.

Traditional approach to $2$-interaction systems usually maps many-body wavefunctions into $2$-particle reduced density matrices ($2$-RMD), which transforms the problem of finding an $N$-particle ground state into a low-dimensional optimization problem \cite{mazziotti2007reduced,coleman2000reduced,mayer1955electron}. However, the image of this mapping has a highly non-trivial geometry, and as we will soon show is dual to the eigenstate moduli problem. The equations required to determine this geometry, known as the $N$-representability conditions, were formulated by Mayer \cite{mayer1955electron} and later Coleman \cite{coleman1963structure}, and became a long-standing problem for many-body quantum systems for more than half a century \cite{harriman1978geometry,erdahl1978representability,erdahl1979two,percus1978role,kryachko2014density}. Despite many recent computational advances \cite{erdahl2000b,nakata2001variational,mazziotti2004realization,zhao2004reduced,cances2006electronic,piris2007natural,verstichel2009variational,shenvi2010active,mazziotti2011large,piris2017global,piris2021global}, exact solutions to this problem either  cast it into other difficult problems \cite{garrod1964reduction,kummer1967n} or use a set of inequalities to approach it \cite{coleman1963structure,garrod1964reduction,erdahl1978representability,zhao2004reduced,fukuda2007large,mazziotti2012structure}. Later examples include Mazziotti's recent development of a complete inequality hierarchy for fermionic 2-RDM representability conditions using tensor decompositions of a set of model Hamiltonians \cite{mazziotti2012structure}. However, apart from a handful of early attempts \cite{kummer1967n, harriman1978geometry}, little is known about the global geometry of the $m$-RDM space.

In this Letter, we address the eigenstate moduli problem. Our approach is inspired by the following observation: The space formed by all pure-state $m$-RDMs is enveloped by a non-trivial boundary that corresponds to the eigenstates of $m$-interaction systems, due to the variational principle. This boundary is formed by all singularities of the mapping from pure states to $m$-RDMs \cite{harriman1978geometry}. Based on this observation, we show that the eigenstate space is a protective variety determined by a set of algebraic equations. This allows us to investigate the global geometry of the eigenstate and RDM spaces. Finally, we generalize our method to the moduli problems for systems with a family of Hamiltonians parameterized by a set of real variables.

We start with a warm-up exercise. Consider a system with Hilbert space $\mathcal{H} = \mathcal{H}_A\otimes \mathcal{H}_B$ divided into two subsystems $A$ and $B$, each with a Hilbert space $\mathcal{H}_A$ and $\mathcal{H}_B$. For simplicity, we assume both are finite-dimensional, i.e,  $\mathcal{H}_A = \mathbb{CP}^{N_A-1}$ and   $\mathcal{H}_B = \mathbb{CP}^{N_B-1}$, and $\mathcal{H} = \mathbb{CP}^{N-1}$ with $N\equiv N_A N_B$.  Without loss of generality, we require $N_A \leq N_B$. Given a pure state $\psi \in \mathcal{H}$, we define a $N_A\times N_B$ rectangular matrix $\Psi \equiv [\psi_{ab}]$, where $ 1\le a\le N_A, 1\le b\le N_B$. The RDM of subsystem $A$, 
\begin{equation}\label{eq:rho0}
\rho^A(\psi, \psi^\dag) \equiv \Psi  \Psi^\dag, 
\end{equation}
projects the pure state $(\psi,\psi^\dag)$ to a Hermitian matrix of subsystem $A$, where $\psi^\dag$ and $\overline\psi$ are the Hermitian conjugate and complex conjugate of $\psi$, respectively. The matrix element of RDM, $\rho_{i,j}^A = \Psi_i^\dag \Psi_j $, where $\Psi_i$ is the transpose of i-th row vector of the matrix $\Psi$. We rewrite $\Psi_i = \hat P_i \psi$ where $\hat P_i = e_A^i$ is the projection operator that projects $\psi$ to $\Psi_i$, and $e_A^i$ is the $N_A$-dimensional row vector with a unit on position $i$. Substituting Eq.~(\ref{eq:rho0}) leads
\begin{equation} \label{eq:rho}
\rho_{i,j}^A = \langle \psi | \hat P_{i,j} | \psi \rangle,
\end{equation}
that maps $\mathcal{H}\times\overline{\mathcal{H}}$ to $\mathcal{D}^A$, where $\overline{\mathcal{H}}$ is the conjugate Hilbert space.  The projection operator
 $\hat P_{i,j} \equiv \hat P_i^t \hat P_j = e^{ij}_A \otimes I_B$, where $e^{ij}_A$ is the matrix with a unit on the $(i,j)$ position and zeros elsewhere, with the trace $\sum_i \hat P_{i,i} =  \hat I_N$ due to the normalization, where $\hat I_N$ is the $N\times N$ identity matrix. 

We are interested in the geometry of the space $\mathcal{D}^A$ formed by all possible RDMs, which is the image of the mapping~(\ref{eq:rho}). In our exercise, $\mathcal{D}^A$ is simply a semialgebraic set consisting of all $N_A \times N_A$ semi-positive definite Hermitian matrices since $\rho_{i,j}^A$ is the inner product between any two vectors $\Psi_i$ and $\Psi_j$. There is a natural stratification of $\mathcal{D}^A$ based on 
the rank of the RDM, $r \equiv \mathrm{rank} (\rho^A) = \mathrm{rank} (\Psi)$, where
each strata $S_r = \partial S_{r+1}$ is the set of all positive semi-definite matrices with rank $r$ that lies in the boundary of the strata $S_{r+1}$.

To investigate the stratified structure of $\mathcal{D}^A$, we next study the singular points of the mapping~(\ref{eq:rho}). Consider the invariant class of infinitesimal changes $\psi \rightarrow \psi + \delta \psi$ that keeps the RDM unchanged. This invariance requires $\delta \rho^A = 0 $, or 
\begin{equation} \label{eq:delta}
   \delta \psi^\dag \hat P_{i,j} \psi + \psi^\dag \hat P_{i,j} \delta \psi = 0.
\end{equation}
Expressing $\delta \psi = \epsilon \hat L \psi$ in terms of an infinitesimal generator $\hat L$ leads
\begin{equation}\label{eq:L0}
\langle \psi | \hat L^\dag \hat P_{i,j} + \hat P_{i,j} \hat L | \psi \rangle = 0.
\end{equation}
The global symmetry of Eq.~(\ref{eq:L0}) corresponds to the operator equation $\hat L^\dag \hat P_{i,j} + \hat P_{i,j} \hat L  = 0$. Tracing out indices leads to $L^\dag + L = 0$,  reflecting $\hat L \in  \mathfrak{su}(N)$. Therefore, all globally invariant generator $\hat L$ forms a Lie subalgebra of $\mathfrak{su}(N)$. Substituting into Eq.~(\ref{eq:L0}) leads
\begin{equation}\label{eq:L}
 [\hat L, \hat P_{i,j}]  = 0,
\end{equation}
from which we obtain $\hat L = \hat I_{N_A} \otimes \hat L_B$, where $\hat L_B \in \mathfrak{su}(N_B)$. Hence, the solution space of $\hat L$ is a diagonal embedding of the Lie subalgebra $\mathfrak{su}(N_B)$ into $\mathfrak{su}(N)$. Indeed, any unitary transformation on the subspace $\mathcal{H}_B$ preserves the inner product $\Psi_i^\dag \Psi_j$, and consequently the reduced density matrix $\rho^A_{i,j}$.

\begin{figure}
  \includegraphics[width=1\linewidth]{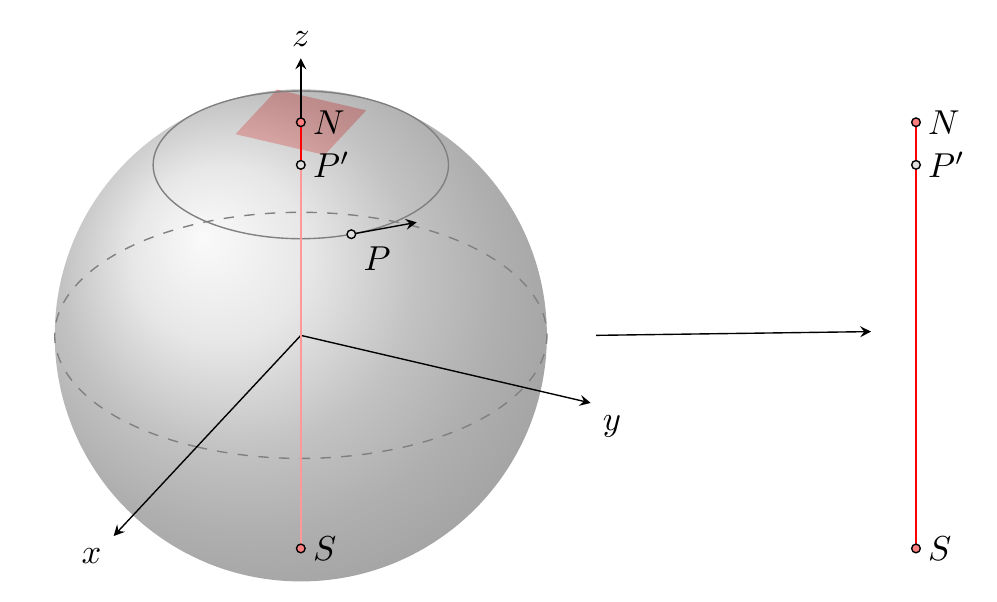}
  \caption{Projection of $S^2$ surface to the line segment $[-1,1]$ on the $z$-axis through $(\theta, \phi)\rightarrow \cos\phi$. The invariant infinitesimal change $0 = \delta z =  -\sin\phi \delta \phi$ requires $\delta \phi = 0$ except for the two poles with $\phi = 0$ or $\pi$. Point $P$, along with points of the same latitude (solid circle), is projected to a single point $P'$. The invariant tangent subspace $\delta \phi = 0$ at point $P$ is one-dimensional, represented by the solid arrow starting from $P$. In contrast, the invariant tangent subspaces of the north and south poles (marked by a red square) are two-dimensional, meaning they map onto the boundaries of the projected space $[-1,1]$.}
  \label{fig:boundary}
\end{figure}

Besides global symmetries, there are non-trivial local symmetries associated with point-dependent $\hat{L}(\psi,\psi^\dag)$ solutions of Eq.~(\ref{eq:L0}). Instead of working on $\hat{L}$ directly, we rewrite Eq.~({\ref{eq:delta}}) as 
\begin{equation} \label{eq:eq:jacobi}
\begingroup
\renewcommand*{\arraystretch}{1.3}
 \mathcal{J}(\psi, \psi^\dag) \begin{bmatrix} \delta \psi \\ \overline{\delta \psi} \end{bmatrix}= 0,
\endgroup
\end{equation}
where $\mathcal{J}(\psi, \psi^\dag) $ is the $N_A^2 \times 2N$ Jacobian matrix of the mapping~({\ref{eq:rho}}), satisfying
\begin{equation} \label{eq:J}
 \mathcal{J}(\psi, \psi^\dag)  = \left [ \psi^{\dag}\hat P_{i,j} ,  \psi^t \hat P_{i,j}^t \right].
\end{equation}
The square bracket represents the row concatenation for all $1\leq i,j \leq N_A$. The kernel of the Jacobian~(\ref{eq:J}) corresponds to the invariant tangent subspace at the point $(\psi, \overline \psi)$, with a dimension $\dim \mathrm{ker} (\mathcal{J}) = 2N - N_A^2$ for generic points. However, singular points that lie in the boundary of $\mathcal{D}^A$, have extra degrees of freedom \cite{hartshorne2013algebraic}, i.e., $\dim \mathrm{ker} (\mathcal{J}) > 2N-N_A^2$. Figure~\ref{fig:boundary} illustrates a toy example of this argument. The index theorem, $\dim \mathrm{ker} ({\mathcal{J}}) - \dim \mathrm{coker} ({\mathcal{J}}) = 2N-N_A^2$ indicates singular points correspond to $\dim \mathrm{coker} ({\mathcal{J}})> 0$. This implies the existence of a non-trivial cokernel of the Jacobian, 
\begin{equation}  \label{eq:coker}
\eta^t \mathcal{J}(\psi, \psi^\dag) = 0,
\end{equation}
with some none-zero $\eta$ for a singular point $(\psi, \psi^\dag)$. We will show later that the cokernel condition~(\ref{eq:coker}) is not only a technical convenience, but has deep physical origins. For our exercise, it can be shown that $\mathrm{coker} (\mathcal{J}) = \mathrm{coker} (\overline \Psi) \times \mathrm{coker} ( \Psi) $, indicating $\dim \mathrm{coker} ({\mathcal{J}}) = \dim \mathrm{coker} ( \Psi)^2 $. Therefore, the preimage of strata $S_r$ in $\mathcal{D}^A$ corresponds to the singularity set with the Jacobian cokernel of the same dimensionality. In particular, the preimage of $\mathcal{D}^A$ boundary corresponds to points satisfying the cokernel condition~(\ref{eq:coker}). For the more general case where $\mathcal{D}^A$ is not directly accessible, its geometry can be studied implicitly using the Jacobian~(\ref{eq:eq:jacobi}) and its cokernel.

We now turn our attention to many-body systems. Much of our discussion in the exercise can be transplanted to this case. Consider a system of $n$ identical particle, each with $q$ states in the single-particle Hilbert space, $\mathcal{H}^1 \equiv \mathbb{CP}^{q-1}$. The Hilbert space of the system  $\mathcal{H}^\pm = \mathrm{Sym}^\pm \otimes^n \mathcal{H}^1$, where $\mathrm{Sym}^\dag$ and $\mathrm{Sym}^-$ are the symmetrization and anti-symmetrization projections, with dimension $N  = \binom{q+n-1}{n}$ and $\binom{q}{n}$ for bosons and fermions, respectively. We further require $2N \ge N_A^2$, which is always valid when $n \gg m$.

Dividing the system into two subsystems $A$ and $B$ with $m$ and $n-m$ particles with $m<n$, we have $\mathcal{H}_A = \mathrm{Sym}^\pm \otimes^k \mathcal{H}^1$ and $\mathcal{H}_B = \mathrm{Sym}^\pm \otimes^{n-m} \mathcal{H}^1$, whose dimensions $N_A = \binom{q+m-1}{m}$ and $N_B =  \binom{q+n-m-1}{n-m}$ for bosons, and $N_A = \binom{q}{m}$ and $N_B = \binom{q}{n-k}$ for fermions. In order to represent the wavefunction in the symmetrized space, we define $\mathcal{I}^\pm_{k} = \{ [i_1, \ldots, i_k],  \forall 1 \leq i_1, \ldots, i_k \leq q \} $ where $ i_1 \leq i_2 \leq \ldots \leq i_k $ for bosons and $ i_1 < i_2 < \ldots < i_k $ for fermions \cite{harriman1978geometry}. An index set $I \in \mathcal{I}^\pm_{k}$ is the same basis as the occupation-number representation. However, the index notation is relatively straightforward for our discussion below. 

The $m$-RDM, $\rho^{(m)}$, defined in Eq.~(\ref{eq:rho}), is associated with bosonic and fermionic projection operators 
\begin{equation}\label{eq:PIJ}
(\hat P_{I,J}^{(m)})_{\alpha\beta} = \sum_K \sigma_{I,K}\sigma^\pm_{J,K}\delta_{\alpha, (IK)}\delta_{\beta, (JK)},
\end{equation}
where indices $I, J \in \mathcal{I}^\pm_m$, $K \in \mathcal{I}^\pm_{n-m}$, and $\alpha, \beta \in \mathcal{I}^\pm_{n}$. The parentheses $(\cdot)$ represent the ordered concatenation of the index set. The symmetry factor for bosons $\sigma^\dag_{I,J} \equiv \sqrt {\frac{(n_1^I!\ldots n_p^I!) (n_1^J!\ldots n_p^J!)}{(n_1^{(IJ)}!\ldots n_p^{(IJ)}!)}}$ counting for the multiplicity from $\mathcal{H}$ to $\mathcal{H}^A\otimes\mathcal{H}^B$, where $n_i^I$ represents the particle number at state $i$ for the index set $I$. The symmetry factor for fermions $\sigma^-_{I,J} = \varepsilon_{I,J}$ is the Levi-Civita tensor that captures the antisymmetric nature of fermions for an unordered index set. The RDM space $\mathcal{D}^{(m)}_{n,q}$ consists of all $\rho^{(m)}$ with real dimension $N_A^2$. Note that one might also use the traditional projection, $\hat P_{i_1,\ldots i_m; j_1,\ldots, j_m} \equiv a_{i_1}\ldots a_{i_m} a_{j_1}^\dag \ldots a_{j_m}^\dag $ that projects $\mathcal{H}$ to the $m$-particle tensor product space $\otimes^m \mathcal{H}^{(1)}$. Equation~(\ref{eq:PIJ}) further reduces it to the $m$-particle occupation number representation. 

Unlike the warm-up exercise where the system's Hilbert space is decomposed into the tensor product of two subsystems, the symmetry of identical particles hinders the decomposability of the projection operator~(\ref{eq:PIJ}). Thus, Equation~(\ref{eq:L}) has a single solution $\hat L = \iu$, corresponding to the trivial $U(1)$ global symmetry. On the other hand, if only $m$-interaction is involved, this extra symmetry allows us to reduce the full Hamiltonian $\hat H \equiv \sum H^{(m)}_{I,J} \hat P_{I,J}^{(m)}$ through $m$-particle Hamiltonian $H^{(m)}$. Therefore, the average energy
\begin{equation}\label{eq:E}
    E= \langle \psi | \hat H |\psi\rangle  = \mathrm{tr}( H^{(m)}\rho^{(m)}).
\end{equation}
The variational principle suggests that the stationary points of Eq.~(\ref{eq:E}) correspond to the eigenstates $\psi$ of $\hat H$, satisfying
\begin{equation}\label{eq:var}
0 =  \sum_{IJ} H^{(m)}_{IJ} \frac{\delta \rho^{(m)}_{I,J}}{\delta \psi} = 
(H^{(m)})^t \mathcal{J},
\end{equation}
which recovers the cokernel condition~(\ref{eq:coker}) with $\eta =  H^{(m)}$. Therefore, the $m$-particle Hamiltonian lies precisely in the cokernel of $\mathcal{ L}$. Geometrically, the eigenstate $\psi$ lies in the preimage of the boundary $\mathcal{M}^{(m)}_{n,q} \equiv \partial \mathcal{D}^{(m)}_{n,q}$, where $ H^{(m)}$ is the normal vector on $\mathcal{M}^{(m)}_{n,q}$ at the point $\rho^{(m)}$. The ground states corresponding to the global minima of Eq.~(\ref{eq:E}) usually form a subspace of this boundary. Since the eigenstates are determined by the normal vector $ H^{(m)}$, there is a one-to-one correspondence between $\psi$ and its RDM image $\rho$ \cite{harriman1978geometry}. Below we will not distinguish the geometry of $\mathcal{M}^{(m)}_{n,q}$ from its preimage, because the point in $\mathcal{M}^{(m)}_{n,q} $ is fully determined by the eigenstates of $m$-interaction systems, and vice versa.

To study $\mathcal{M}^{(m)}_{n,q}$, we requires the Jacobian for infinitesimal changes. Substituting (\ref{eq:J}) with (\ref{eq:PIJ}) obtains
\begin{equation}\label{eq:J1}
\mathcal{J}^{(m)}_{IJ, (IK)} =  \left[\sigma_{I,K}\sigma_{J,K}\overline\psi_{(IK)}, \sigma_{I,K}\sigma_{J,K}\psi_{(JK)}\right],
\end{equation}
and zeros elsewhere. The cokernel condition~(\ref{eq:coker})
is equivalent to the vanishing of all $N_A^2\times N_A^2$ minors, as 
\begin{equation}\label{eq:minor}
\det M_{I_1,\ldots, I_k, J_1, \ldots, J_{k'}}  = 0
\end{equation}
where $ M_{I_1,\ldots, I_k, J_1, \ldots, J_{k'}}$ represents the submatrix formed by selecting $\{I_1,\ldots, I_k \}$-th $\psi$ columns and $\{J_1,\ldots, J_{k'} \}$-th $\overline\psi$ columns in Eq.~(\ref{eq:J1}), with $k+k' = N_A^2$ and $k,k' \leq N$.

The minor condition~({\ref{eq:minor}) generates a homogeneous ideal $I^{(m)}_{n,q}$ over the polynomial ring $K[\{\psi\},\{\overline{\psi}\}]$. Consequently, it determines the eigenstate spaces  $\mathcal{M}^{(m)}_{n,q} \equiv \mathbb{C}[\{\psi\},\{\overline{\psi}\}]/I^{(m)}_{n,q}$ as the corresponding homogeneous coordinate ring. Moreover, $\rho^{(m)}$ is a tensor contraction of $\rho^{(m')}$  for any $m' > m$, together with Eqs.~(\ref{eq:rho}) \& (\ref{eq:J}) implying that $\mathcal{J}^{(m)}$ spans a linear subspace of $\mathcal{J}^{(m')}$. Consequently, if all minors for $\mathcal{J}^{(m)}$ in Eq.~(\ref{eq:minor}) vanish they also vanish for any $m' > m$, leading to a descending chain of ideals,
\begin{equation}
I^{(1)}_{n,q} \supset I^{(2)}_{n,q} \ldots  \supset I^{(n)}_{n,q} = 1. 
\end{equation}
This suggests a filtration of the corresponding eigenstate spaces,
\begin{equation}
\mathcal{M}^{(1)}_{n,q} \subset \mathcal{M}^{(2)}_{n,q} \ldots \subset \mathcal{M}^{(n)}_{n,q} = \mathcal{H}\times\overline{\mathcal{H}},
\end{equation}
reflecting the fact that the $m'$-particle interacting systems are special cases for $m'$-particle systems with $m < m'$.

To better understand these ideals, we first investigate the property of $I^{(1)}_{n,q}$, since all higher-order ideals $I^{(m)}_{n,q}$ for $m>1$ are its subideals. For fermions, the $m=1$ eigenstates are characterized by $n$ Hermitian orthogonal 1-particle wavefunctions $\{\psi^{(1)}\ldots\psi^{(n)} \} \in  \mathcal{H}^1$. The Pl\"ucker embedding \cite{hartshorne2013algebraic} maps these 1-particle wavefunctions to the Slater determinant $\psi = \psi^{(1)}_1 \wedge \ldots \wedge \psi^{(1)}_n$. Consequently, the space formed by all 1-interaction fermionic eigenstates is the unitary Grassmannian $\mathsf{UG}(n,\mathcal{H}^{1}\times \overline{\mathcal{H}^{1}}) \subset \mathsf{Gr}(n,\mathcal{H}^{1}) \times \mathsf{Gr}(n,\overline{\mathcal{H}^{1})}$. These eigenstates $\psi$ satisfy a set of quadratic Pl\"ucker relations, 
\begin{equation}\label{eq:Plucker}
\sum_{i \in J} (-1)^{\mathrm{ord}(i)} \psi_{[Ii]} \psi_{[J\slashed{i}]} = 0,
\end{equation}
where $I \in \mathcal{I}_{n-1}$, $J \in \mathcal{I}_{n+1}$ and $\slashed{i}$ represent the removal of $i$ from the index set $J$. For instance, for $n=2$ and $p = 4$ one has a Skein-like relation $\psi_{12}\psi_{34} - \psi_{13}\psi_{24} + \psi_{14}\psi_{23} = 0$, as illustrated in Fig.~{\ref{fig:pluker}}. Moreover, the Hermitian orthogonality requires 
\begin{equation} \label{eq:Hermite}
\psi_{I}^\dag \psi_{J} = 0,
\end{equation}
for all $I \neq J$. Therefore, the ideal $I^{(1)}_{n,q}$ of fermions is generated by all Pl\"ucker relations~(\ref{eq:Plucker}) (and their conjugates), and Hermitian orthogonality~(\ref{eq:Hermite}). Since the unitary Grassmannian is smooth, $\mathcal{D}^{(1)}_{n,q}$ has no other strata. The ground state space is also equal to $\mathcal{M}^{(1)}_{n,q}$, because the 1-particle Hamiltonian can be smoothly deformed to exchange the ground and excited states.  

Similarly, the bosonic eigenstates $\psi = \psi^{(1)} \vee \ldots \vee \psi^{(n)}$ are the permanents of single-particle wavefunctions, which defines the $n$-uple symmetric Segre embedding \cite{hartshorne2013algebraic}. Unlikely the fermionic case, these single-particle wavefunctions are not necessarily independent. Denoting $r$ as the rank of these functions, there is a natural stratification, where each strata $S_r$ is classified by a unique cokernel dimension $\dim \mathrm{coker}(\mathcal{J}) = d^{r}_{n,q}$ deceasing monotonically with $r$ for fixed $q,n$. The explicit form of $d^{r}_{n,q}$ is unknown except for the lowest $r=1$ we have $d^{1}_{n,q} = (q-1)^2$. This the is the case for the ground state consisted of factorizable wavefunctions, $\psi_{i_1,\ldots,i_n} = \prod_{k=1}^n \psi^{(1)}_{i_k}$, characterized by a single 1-particle $\psi^{(1)}$. They form a smooth subvariety of $\mathcal{M}^{(1)}_{n,q}$, which is known as the Veronese variety. Up to a normalization factor, the Veronese variety also satisfies a set of quadratic relations,
\begin{equation}\label{eq:Vernose}
\psi_{I} \psi_{J} - \psi_{I'} \psi_{J'} = 0,
\end{equation}
for all $(IJ) = (I'J')$. They generate the ideal $\langle \psi_{I} \psi_{J}- \psi_{I'} \psi_{J'},  \overline\psi_{I} \overline\psi_{J}- \overline\psi_{I'} \overline\psi_{J'}\rangle$ for all 1-particle ground states. 
\begin{figure}
  \includegraphics[width=1\linewidth]{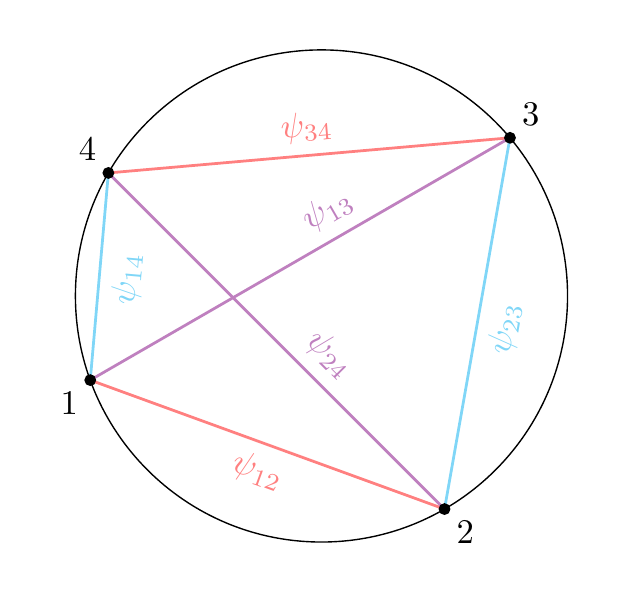}
  \caption{The Pl\"ucker relations of $n=2$ and $p = 4$ can be visualized as Ptolemy's theorem for an inscribable quadrilateral on a circle, $\psi_{13}\psi_{24} = \psi_{12}\psi_{34} + \psi_{14}\psi_{23} $.  }
  \label{fig:pluker}
\end{figure}

What about $m = 2$, the most physically relevant case? To validate the cokernel condition~(\ref{eq:coker}), we perform numerical simulations of bosonic systems for $q=2$ and $q = 3$ with $n=4$, $6$ and $8$ particles. We randomly generate a set of pairwise Hamiltonians $H^{(2)}$ and extract all eigenstates numerically using the exact diagonalization method \cite{weisse2008exact}. We compute the cokernel dimension of Eq.~(\ref{eq:J1}) for eigenstates. Our numerical results show that $\dim \mathrm{coker}(\mathcal{J}) = 1$ for all eigenstates generated in our simulation. We further verify the minor conditions numerically in a set of randomly selected submatrices. In contrast, we generate a set of random $\psi$ and find that their $\mathcal{J}^{(2)}$ have no non-trivial cokernels. Similar to $m=1$, we also find the bosonic $\mathcal{M}^{(2)}$ has a non-trivial stratification classified by the cokernel dimension of Jacobian. Apart from the full minor condition~(\ref{eq:coker}), little is known analytically. Nonetheless, the fact $I^{(2)}_{n,q}$ is a subideal of $I^{(1)}_{n,q}$ implies that their generators are polynomials in $I^{(1)}_{n,q}$. Since the higher order varieties $I^{m}_{n,q}$ can be viewed as generalizations of the Grassmannian and the Veronese variety, a simple set of generators similar to Eqs.~(\ref{eq:Plucker} -- \ref{eq:Vernose}) is expected. A comprehensive answer to this question, including finding the minimal free resolution and syzygy \cite{weyman2003cohomology}, is beyond the scope of this Letter. 

In general, the minor condition~(\ref{eq:minor}) completely determines the boundary geometry of the space $\mathcal{D}^{(m)}$ and implicitly determines the pure-state $N$-representability conditions. Expressing the representability conditions in terms of $\rho_{m}$ requires transforming the homogeneous coordinates $\psi$ into affine coordinates $\rho_{m}$. The Zariski closure of the RDM boundary can be formally written as $\langle I^{(m)}(\psi, \overline\psi ), \psi^\dag \hat P_{I,J} \psi  - \rho^{(m)}_{IJ}\rangle \cap \mathbb{C}[\rho^{(m)}_{IJ}] $, by substituting $\rho^{(m)}$ and canceling $\psi$. Since the boundary for $m > 1$ has a codimension one, this transformation leaves only one algebraic equation for $\rho^{(m)}$. For small $(n,q)$ systems, this equation may be derived explicitly by computer. While the $N$-representability conditions are known to be NP-hard \cite{deza1997geometry} in general, it is still very interesting to seek for this algebraic equation in future studies.

Our approach can be further generalized to the systems with a family of Hamiltonians parameterized by a set of real variables. Consider the Schr\"odinger equation 
\begin{equation}\label{eq:schrodinger}
    (\hat H(\eta) - E) \psi = 0,
\end{equation}
for a family of Hamiltonians $\hat H = \sum_{\alpha} \eta_\alpha \hat H_\alpha$ as a linear combination of a set of Hermitian operators $H_\alpha$ and real parameters $\eta_\alpha$. The textbook approach to Eq.~(\ref{eq:schrodinger}) usually looks for the eigenstate $\psi$ of a fixed  parameter set $\eta$. Here we consider the dual problem, that is, solving $\eta$ for a fixed $\psi$.  We rewrite the Schr\"odinger equation~({\ref{eq:schrodinger}}) as 
\begin{equation}
\eta^t \mathbf{\hat H} \psi = 0, 
\end{equation}
 where $\mathbf{\hat H} \equiv \{\hat H_\alpha\}$ and  $\eta \equiv \{\eta_\alpha\}$ also include $\hat H_0 \equiv \hat I$ and $\eta_0 \equiv -E$. Together with its conjugate since the parameters are real, we recover the cokernel condition~(\ref{eq:coker}), where 
\begin{equation}\label{eq:Lall}
 \mathcal{J}(\psi, \psi^\dag)  \equiv \left [ \psi^{\dag}\hat H_\alpha  ,  \psi^t \hat H_\alpha ^t \right],
\end{equation}
 is a $N_p \times 2N$ rectangular matrix where $\eta \in \mathbb{RP}^{N_p-1}$ with $N_p$ number of parameters. The parameter set $\eta^t \in \mathrm{coker} (\mathcal{J})$ lies precisely in the cokernel of $ \mathcal{J}$, reflecting that the cokernel condition~(\ref{eq:coker}) is dual to the Schr\"odinger equation~(\ref{eq:schrodinger}). The minor condition~(\ref{eq:minor}) completely determines the geometry of the eigenstate space of a quantum family parameterized by $\eta$. 
 
 To apply Eq.~(\ref{eq:Lall}) to the $m$-interaction systems, we introduce $\hat H_{\alpha \equiv (I,J) } \equiv (\hat P_{I,J}+\hat P_{J,I}) +  \iu (\hat P_{I,J} - \hat P_{J,I}) $ that encodes $N_A^2$ real operators $\hat P_{I,J}$ into Hermitian operators $\hat H_{I,J}$. Similarly, we encode the $m$-particle Hamiltonian $\hat H^{(m)}_{I,J} = (\eta_{I,J}+\eta_{J,I}) +  \iu (\eta_{I,J} - \eta_{J,I}) $. It is not difficult to verify that the original Hamiltonian $\hat H \equiv \sum_{I,J} H^{(m)}_{I,J} \hat P_{I,J} = \sum_{I,J} \eta_{I,J} \hat H_{I,J}$, or equivalently $( H^{(m)})^t \mathcal{J}' = \eta^t \mathcal{J}$, where $\mathcal{J}'$ is precisely the one defined in Eq.~(\ref{eq:J}). This result reveals again the key finding of our paper~(\ref{eq:var}), that the $m$-Hamiltonian lies in the cokernel of $\mathcal{J}$ and connects our theory to existing approaches to the $N$-representability conditions using a set of model Hamiltonians \cite{mazziotti2012structure}.
 
 Equation~(\ref{eq:Lall}) provides a general solution to the moduli problem. Taking the Hubbard model as example, its Hamiltonian $\hat H = t \hat T +  U \hat V$,
where $\hat T \equiv \sum_{\langle ij \rangle,\sigma}  \hat c_{i,\sigma}^\dag \hat c_{j,\sigma} + \hat c_{j,\sigma}^\dag \hat c_{i,\sigma} $ and $\hat V\equiv \sum_i  \hat n_{i \uparrow} \hat n_{i\downarrow}$. Substituting $\eta = \{-E, t, U\}$ and $\mathbf{\hat H}  =\{\hat I, \hat T, \hat V\} $ into  Eq.~(\ref{eq:Lall}) leads $ \mathcal{J}_\mathrm{Hubbard}  = 
  \begin{pmatrix}
   \psi,  & 
  \hat T \psi, & 
  \hat V \psi 
  \end{pmatrix}^t$.
Here we discard the redundant conjugate part since $\hat T$, $\hat V$ and $\psi$ are real numbers. Because the minor condition only involves $3\times3$ submatrices, the eigenstates of the Hubbard model form a real cubic variety. While solving the minor condition directly seems as difficult as solving the Schr\"odinger equation~(\ref{eq:schrodinger}), one may wish to extract interesting geometry information encoded in this variety, especially when the system undergoes a phase transition. We will leave these possibilities to future investigations.

In conclusion, our results provide a comprehensive approach to the eigenstate moduli problem and offer a global geometric picture to the pure-state $N$-representability conditions. Mixed-state RDM space can be viewed as a chordal variety of pure-state RDMs. It also poses many challenges for future research. One possible direction is to understand better the connection between our method and  existing approaches \cite{mazziotti2012structure}, which likely connects to the stratified structure of the RDM space. More generally, one might ask how much useful information for strongly correlated systems can be extracted from the algebraic equations we have discovered. Answering this question requires bringing modern mathematical tools from algebraic geometry \cite{gharahi2020fine}. On the other hand, the eigenstate varieties generalize many classical algebro-geometric objects including the Grassmannian and the Veronese variety, which is interesting in its own right. Our approach provides a new perspective on many-body quantum systems, with potential implications for strongly correlated systems, entanglement measurements, and quantum computational chemistry.  



%

\end{document}